\documentstyle[12pt]{article}
\input epsf.tex
\begin{document}
\catcode`@=11
\newif\if@defeqnsw \@defeqnswtrue
 
\def\eqnarray{\stepcounter{equation}\let\@currentlabel=\theequation
\if@defeqnsw\global\@eqnswtrue\else\global\@eqnswfalse\fi
\global\@eqnswtrue
\tabskip\@centering\let\\=\@eqncr
$$\halign to \displaywidth\bgroup\hfil\global\@eqcnt\z@
  $\displaystyle\tabskip\z@{##}$&\global\@eqcnt\@ne
  \hfil$\displaystyle{{}##{}}$\hfil
  &\global\@eqcnt\tw@ $\displaystyle{##}$\hfil
  \tabskip\@centering&\llap{##}\tabskip\z@\cr}
 
\def\yesnumber{\global\@eqnswtrue}
 
\def\@@eqncr{\let\@tempa\relax\global\advance\@eqcnt by \@ne
    \ifcase\@eqcnt \def\@tempa{& & & &}\or \def\@tempa{& & &}\or
     \def\@tempa{& &}\or \def\@tempa{&}\else\fi
     \@tempa \if@eqnsw\@eqnnum\stepcounter{equation}\fi
     \if@defeqnsw\global\@eqnswtrue\else\global\@eqnswfalse\fi
     \global\@eqcnt\z@\cr}
 
 
\def\@eqnacr{{\ifnum0=`}\fi\@ifstar{\@yeqnacr}{\@yeqnacr}}
 
\def\@yeqnacr{\@ifnextchar [{\@xeqnacr}{\@xeqnacr[\z@]}}
 
\def\@xeqnacr[#1]{\ifnum0=`{\fi}\cr \noalign{\vskip\jot\vskip #1\relax}}
 
\def\eqalign{\null\,\vcenter\bgroup\openup1\jot \m@th \let\\=\@eqnacr
\ialign\bgroup\strut
\hfil$\displaystyle{##}$&$\displaystyle{{}##}$\hfil\crcr}
\def\endeqalign{\crcr\egroup\egroup\,}
 
 
\def\cases{\left\{\,\vcenter\bgroup\normalbaselines\m@th \let\\=\@eqnacr
    \ialign\bgroup$##\hfil$&\quad##\hfil\crcr}
\def\endcases{\crcr\egroup\egroup\right.}
 
 
\def\eqalignno{\stepcounter{equation}\let\@currentlabel=\theequation
\if@defeqnsw\global\@eqnswtrue\else\global\@eqnswfalse\fi
\let\\=\@eqncr
$$\displ@y \tabskip\@centering \halign to \displaywidth\bgroup
  \global\@eqcnt\@ne\hfil
  $\@lign\displaystyle{##}$\tabskip\z@skip&\global\@eqcnt\tw@
  $\@lign\displaystyle{{}##}$\hfil\tabskip\@centering&
  \llap{\@lign##}\tabskip\z@skip\crcr}
 
\def\endeqalignno{\@@eqncr\egroup
      \global\advance\c@equation\m@ne$$\global\@ignoretrue}

 
\@namedef{eqalignno*}{\@defeqnswfalse\eqalignno}
\@namedef{endeqalignno*}{\endeqalignno}
 
 
\def\eqaligntwo{\stepcounter{equation}\let\@currentlabel=\theequation
\if@defeqnsw\global\@eqnswtrue\else\global\@eqnswfalse\fi
\let\\=\@eqncr
$$\displ@y \tabskip\@centering \halign to \displaywidth\bgroup
  \global\@eqcnt\m@ne\hfil
  $\@lign\displaystyle{##}$\tabskip\z@skip&\global\@eqcnt\z@
  $\@lign\displaystyle{{}##}$\hfil\qquad&\global\@eqcnt\@ne
  \hfil$\@lign\displaystyle{##}$&\global\@eqcnt\tw@
  $\@lign\displaystyle{{}##}$\hfil\tabskip\@centering&
  \llap{\@lign##}\tabskip\z@skip\crcr}
 
\def\endeqaligntwo{\@@eqncr\egroup
      \global\advance\c@equation\m@ne$$\global\@ignoretrue}
 
\@namedef{eqaligntwo*}{\@defeqnswfalse\eqaligntwo}
\@namedef{endeqaligntwo*}{\endeqaligntwo}
 
%
%
%
%
%
\newtoks\@stequation
 
\def\subequations{\refstepcounter{equation}%
  \edef\@savedequation{\the\c@equation}%
  \@stequation=\expandafter{\theequation}
  \edef\@savedtheequation{\the\@stequation}
  \edef\oldtheequation{\theequation}%
  \setcounter{equation}{0}%
  \def\theequation{\oldtheequation\alph{equation}}}
 
\def\endsubequations{%
  \setcounter{equation}{\@savedequation}%
  \@stequation=\expandafter{\@savedtheequation}%
  \edef\theequation{\the\@stequation}%
  \global\@ignoretrue}
 
 
\def\big#1{{\hbox{$\left#1\vcenter to1.428\ht\strutbox{}\right.\n@space$}}}
\def\Big#1{{\hbox{$\left#1\vcenter to2.142\ht\strutbox{}\right.\n@space$}}}
\def\bigg#1{{\hbox{$\left#1\vcenter to2.857\ht\strutbox{}\right.\n@space$}}}
\def\Bigg#1{{\hbox{$\left#1\vcenter to3.571\ht\strutbox{}\right.\n@space$}}}
 
\catcode`@=12
%
%
\newcommand{\gsimeq}{\mathop{>}\limits_{\displaystyle{\sim}}}
\newcommand{\lsimeq}{\mathop{<}\limits_{\displaystyle{\sim}}}
\newcommand{\sigmin}{\sigma_{\gamma \gamma}^{\rm min}}
\newcommand{\as}{\alpha_s}
\newcommand{\amu}{A_t + \mu \cot \! \beta}
\newcommand{\mgg}{M_{\gamma \gamma}}
\newcommand{\fourl}{l^+ l^+ l^- l^-}
\newcommand{\fourtau} {\tau^+ \tau^+ \tau^- \tau^-}
\newcommand{\tht}{\theta_t}
\newcommand{\tanb}{\tan \! \beta}
\newcommand{\msb}{m_{\tilde{b}_L}}
\newcommand{\msbsq}{m^2_{\tilde{b}_L}}
\newcommand{\mt}{m_t}
\newcommand{\mtsq}{m^2_t$}
\newcommand{\mst}{m_{\tilde{\tau}}}
\newcommand{\mse}{m_{\tilde{e}}}
\newcommand{\msmu}{m_{\tilde{\mu}}}
\newcommand{\msti}{m_{\tilde{\tau}_1}}
\newcommand{\mstii}{m_{\tilde{\tau}_2}}
\newcommand{\mstl}{m_{\tilde{\tau}_L}}
\newcommand{\mstr} {m_{\tilde{\tau}_R}}
\newcommand{\msl} {m_{\tilde{l}}}
\newcommand{\mstlsq}{m^2_{\tilde{\tau}_L}}
\newcommand{\mstrsq}{ m^2_{\tilde{\tau}_R}}
\newcommand{\mstlr}{m_{\tilde{\tau}_{L,R}}}
\newcommand{\mstlrsq}{ m^2_{\tilde{\tau}_{L,R}}}
\newcommand{\mstssq}{ m^2_{\tilde{\tau}_1}}
\newcommand{\msttsq}{ m^2_{\tilde{\tau}_2}}
\newcommand{\stst}{\tilde{t}_1 \tilde{t}_1^*}
\newcommand{\sigst} { \sigma_{\tilde{t}_1} }
\newcommand{\msig} { m_{\sigma_{\tilde t}} }
\newcommand{\gamgam} { \gamma \gamma }
\newcommand{\selr} { \tilde{e}_{L,R} }
\newcommand{\smulr} { \tilde{\mu}_{L,R} }
\newcommand{\se} { \tilde{e}}
\newcommand{\smu} { \tilde{\mu}}
\newcommand{\sti} { \tilde{\tau}_1 }
\newcommand{\stii} { \tilde{\tau}_2 }
\newcommand{\thetat}{\theta_{\tau}}
\newcommand{\cost}{\cos\theta_{\tau}}
\newcommand{\costs}{\cos^2\theta_{\tau}}
\newcommand{\sint}{\sin\theta_{\tau}}
\newcommand{\sints}{\sin^2\theta_{\tau}}
\newcommand{\sw}{\sin\!\theta_W}
\newcommand{\cw}{\cos\!\theta_W}
\newcommand{\tw}{\tan\!\theta_W}
\newcommand{\brchi}{Br(\sti\rightarrow \chi_1^0 \tau)}
\newcommand{\gami}{\Gamma(\sti\rightarrow\chi_1^0\tau)}
\newcommand{\gamii}{\Gamma(\sti\rightarrow\chi_2^0\tau)}
\newcommand{\gamci}{\Gamma(\sti\rightarrow\chi_1^-\tau)}
\newcommand{\cosb}{\cos\!\beta}
\newcommand{\sinb}{\sin\!\beta}
\newcommand{\sws}{\sin^2\theta_W}
\newcommand{\cws}{\cos^2\theta_W}
\newcommand{\tws}{\tan^2\theta_W}
\newcommand{\mtau}{m_{\tau}}
\newcommand{\chio}{ \chi^0_1}
\newcommand{\chii}{\chi^0_i}
\newcommand{\chimi}{\chi^-_i}
\newcommand{\aiiR}{a_{1i}^R}
\newcommand{\aiiL}{a_{1i}^L}
\newcommand{\st}{\tilde{\tau}}
\newcommand{\stl}{ \tilde{\tau}_L}
\newcommand{\str}{ \tilde{\tau}_R}
\newcommand{\mchi}{m_{{\chi}^0_1}}
\newcommand{\mchip}{m_{{\chi}^+_1}}
\newcommand{\mchisi}{m_{{\chi}^0_i}}
\newcommand{\mchii}{m_{{\chi}^0_2}}
\newcommand{\ptau}{ P_{\tau}}
\newcommand{\ww}{ W^+W^- }
\newcommand{\epem}{ e^+e^-}
\newcommand{\mat}{{\cal M}^2_{\tilde{t}}}
\newcommand{\etauc}{E_{\tau}^{cm}}
\newcommand{\ptauc}{p_{\tau}^{cm}}
\newcommand{\vstau}{v_{\sti}}
\newcommand{\etaucs}{E_{\tau}^{cm 2}}
\newcommand{\etaumax}{E_{\tau}^{\rm max}}
\newcommand{\etaumin}{E_{\tau}^{\rm min}}
\newcommand{\be}{\begin{equation}}
\newcommand{\ee}{\end{equation}}
\newcommand{\een}{\end{subequations}}
\newcommand{\ben}{\begin{subequations}}
\newcommand{\beq}{\begin{eqalignno}}
\newcommand{\eeq}{\end{eqalignno}}
\renewcommand{\thefootnote}{\fnsymbol{footnote} }
\noindent
\begin{flushright}
KEK--TH--435\\
KEK Preprint 95--12\\
April 1995
\end{flushright}
\vspace{1.5cm}
\pagestyle{empty}
\begin{center}
{\Large \bf 
Precision Study of Minimal Supersymmetric Standard Model 
by production and decay of scalar tau lepton
}\\
\vspace*{5mm}
Mihoko M. Nojiri\footnote{E--mail: nojirim@theory.kek.jp. 
talk at 
the 5th workshop on Japan Linear Collider (JLC) at Kawatabi (Feb. 16-17)}\\
{\em Theory Group, KEK, Oho 1--1, Tsukuba, Ibaraki 305, Japan. }
\end{center}
 
\begin{abstract}
Study of the production and decay of scalar tau lepton ($\st$) 
at future $\epem$ colliders helps to determine 
the value of $\tan\beta$ through the measurement of 
the polarization of $\tau$ lepton that arises from $\st$ decay.
Key maps of the parameter space of MSSM are presented.
\end{abstract}
\setcounter{footnote}{0}
\pagestyle{plain}
\section*{1) Introduction}
 
The Minimal Supersymmetric Standard Model (MSSM)\cite{1} is one of the most
promising candidates of the models beyond the Standard Model (SM). It
predicts the existence of superpartners of SM particles below a few TeV to
remove quadratic divergences which appear in radiative corrections 
of the SM Higgs sector;
thus the model is free from the so--called hierarchy problem of GUT
models. It should be noted that the gauge couplings unify very precisely at
high energy scale in MSSM \cite{2}, consistent with SUSY 
SU(5) GUT predictions. 
 
Thus searches of  SUSY particles at future $e^+e^-$ colliders would be one of
its important physics targets. Furthermore, a highly polarized electron beam
available for the future linear colliders reduces the background from $W^+W^-$
pair production to the SUSY signals drastically, making  it possible to study
SUSY parameters very precisely \cite{3}. It was also demonstrated that  some
SUSY parameters, such as masses and couplings of SUSY particles  can be
measured very precisely by studying the production and decay of the first and
second generation of sleptons ($\tilde{e},\tilde{\mu}$) and the lighter
chargino ($\chi^+_1$)\cite{3,3a}. The precise measurements of those parameters
would severely constrain supergravity and superstring models, which predict
relations between various soft SUSY breaking parameters at the Planck
scale\cite{4}. 
 
 
In this talk, I would like to report on  a new  study of 
the production and the decay of 
the scalar tau ($\st$). 
This channel turns out to contain  novel information about the 
tau Yukawa coupling $Y_{\tau}$ or $\tanb$\cite{5}, which is 
very difficult to determine by studying other modes.
 
$\st$ production and decay is different from that of $\tilde{e}$ and
$\tilde{\mu}$ because the (scalar) tau lepton has a 
non--negligible Yukawa coupling $Y_{\tau}\propto \mtau/\cosb$ ; the
coupling would be enhanced linearly $\propto \tan\beta$ 
for large value of $\tan\beta$. 
 
A consequence of the non-negligible 
Yukawa coupling is existence of left-right mixing of $\st$. 
The lighter mass eigenstate of $\st$ would be lighter than 
$\tilde{e}$ or $\tilde{\mu}$, even if mass parameter of 
$\st$ is equal to that of $\tilde{e}$ and $\tilde{\mu}$. 
This will be discussed briefly in section 2. 
 
The same Yukawa coupling appears as
a non--negligible $\tau \st \tilde{H_1^0}$ coupling, where $\tilde{H_1^0}$ is
a neutral higgsino. This interaction is involved in $\st$ decay
into a neutralino $\chi^0_i$ and $\tau$, since the $\chi$'s are 
mixtures of higgsinos and gauginos. Another
feature of $\st$ decay that distinguishes it from other slepton decays is that
the $\tau$ lepton arising from  the decay $\st\rightarrow\chi^0_i\tau$  decays
further in the detector, which enables us to measure the average polarization
of the $\tau$ ($\ptau$) \cite{6,7,8}. 
One can then determine a combination of  the 
higgsino--gaugino mixing of $\chi_i^0$ {\em and } $\tanb$ 
by measuring both the cross section for
$\st$ production and the $\ptau(\st\rightarrow\chii\tau)$. 
Especially the sensitivity of $\ptau$ to $\tan\beta$ 
helps us to determine  $\tanb(> 5)$, by combining the 
information from the other mode. In section 3 and 4 we are going to discuss 
this in some detail.
 
\section*{2) The Model}
To be more specific, we describe the SUSY parameters that appear in 
the MSSM. In this model, the Higgs sector consists of two SU(2) doublets, 
$H_1$ and $H_2$,
and the coupling to the matter sector is described by the superpotential 
\be
W= Y_l H_1\cdot L E^c  + Y_d H_1\cdot Q D^c  + Y_u H_2\cdot QU^c  .
\ee
Here $E$, $D$, and $U$ are $SU(2)$ singlet lepton and quark superfields, 
while  $L$ and $ Q$ are SU(2) doublet sfermion superfields respectively.  
Both of the neutral components of Higgs doublets ($H_1^0$, $H_2^0$) 
would have vacuum expectation 
values and we define $\tan\beta=\langle H_1^0\rangle /\langle H_2^0\rangle$.  
 Yukawa couplings $Y$ are related to $\beta$ as  
$Y_{\tau (b)}= g m_{\tau (b)}/(\sqrt{2}m_W \cosb)$
and $Y_{t}=g m_t/(\sqrt{2}m_W \sinb)$  respectively. It should 
be noted that $Y_{\tau (b)}$ is not negligible for large value of 
$\tanb$. 
 
Superpartners of higgsinos and gauginos mix due to $SU(2)\times U(1)$
symmetry breaking. Its neutral and charged mass eigenstates 
are called neutralinos $\chi_i^0 (i=1, 2, 3, 4)$ and 
charginos $\chi^+_i(i=1,2)$, and their mass matrices are 
described as follows;
 
\ben\label{e9}
\beq
{\cal M}_N&(\tilde{B},\tilde{W_3},\tilde{H_1^0},\tilde{H_2^0})=
\nonumber\\
&\left(\begin{array}{cccc}
M_1 &0&-m_Z\sw\cosb&m_Z\sw\sinb\\
0 &M_2& m_Z\cw\cosb&-m_Z\cw\sinb\\
-m_Z\sw\cosb&m_Z\cw\cosb&0&-\mu \\
m_Z\sw\sinb&-m_Z\cw\sinb&-\mu & 0
\end{array}\right),\nonumber\\
\label{e9a}
\\
{\cal M}_C&(\tilde{W},\tilde{H})=\left(\begin{array}{cc}
M_2 & m_W \sqrt{2}\sinb\\
m_W\sqrt{2}\cosb & \mu
\end{array}
\right).
\label{e9b}
\eeq
\een
Here $M_1$ and $M_2$ are soft breaking gaugino mass parameters of $\tilde B$ 
and $\tilde W$, while $\mu$ is a supersymmetric Higgsino mass parameter. 
These mass matrices are diagonalized by a real orthogonal matrix $N$ for ${\cal
M}_N$, and unitary matrices $U$ and $V$ for ${\cal M}_C$ as follows: 
\be
U^*{\cal M}_C V^{-1}=M_D^C, \ \ N {\cal M}_N N^{-1}=M_D^N. 
\label{e10}
\ee

Due to the  R-parity conservation of MSSM and some cosmological constraint, 
the lightest neutralino $\chi^0_1$ is likely the lightest SUSY particle 
(LSP)and  stable, 
thus escapes from detection at collider experiments. We assume this
throughout the discussions of this paper.

 Left and right scalar fermions also mix due to the $SU(2)\times U(1)$ 
symmetry breaking. However, the mixing is negligible  
for the first and the second generation sfermions. 
The  mass matrix of scalar tau lepton
 $\tilde{\tau}_{L(R)}$  would be described as 
\be\label{e1}
{\cal M}^2_{\st}=\left(\begin{array}{cc}m_{LL}^2 & m_{LR}^2\\
m_{LR}^2& m_{RR}^2\end{array}\right)
=\left( \begin{array}{cc}
m_L^2 + m_{\tau}^2 + 0.27 D &   -m_{\tau}(A_{\tau} + \mu \tan\beta)\\
 -m_{\tau}(A_\tau + \mu \tan\beta)&  m_R^2 +m_{\tau}^2 + 0.23D\\
 \end{array}\right).
\begin{array}{c}\stl\\\str\end{array}
\ee
where $m_R $ and $m_L$ are soft breaking scalar mass parameters of 
$\str$ and $(\tilde{\nu}_{\tau},\st)_L$, $A_{\tau}$ is a trilinear coupling of
$\st_L\st_R H_1$ and   $D=-m_Z^2
\cos(2\beta)$. $\str$ and $\stl$ then mix to form two mass eigenstates $\sti$
and $\stii$ ($\msti<\mstii$) 
\be\label{e2}
\left(\begin{array}{c} \sti\\\stii\end{array}\right)
=\left(\begin{array}{cc}\cost &\sint\\
 -\sint&\cost\end{array}\right)
\left(\begin{array}{c} \stl\\\str\end{array}\right).
\ee
 
Notice 
$m_{LR}^2$ could be  large for very large 
value of $\tanb(\mu)$, so that $m_{\tilde \tau_1}$ is
smaller than $m_{LL}$ or $m_{RR}$.
In the models which predict the equal scalar masses at GUT scale such
as the minimal supergravity model or the superstring model, 
$\mst$ can  be lighter than $\mse$ or $\msmu$. 
This is because of  the $\st$ mixing 
and also the effect of the negative RG running of $m_{LL(RR)}$
of $\st$ 
by $\tau$ Yukawa coupling which makes the mass parameter smaller than 
those of $\tilde{e}$ and $\tilde{\mu}$ at the weak scale.
$\st$ analysis is important in the sense that  it might be found earlier 
than the other sfermions in future collider
experiments.
 
\section*{3) $\st$ decay }
It was demonstrated in Ref.\cite{3} that some of the 
above mass parameters can be determined precisely by  
proposed linear colliders with a highly polarized electron beam. 
The masses  of the lightest neutralino, the lighter chargino, the 
selectron and the smuon were shown to be determined with an  
error of  a few \% for a representative parameter set
by the energy distribution of leptons or jets 
coming from decaying sparticles.
Furthermore, by measuring other quantities such as the production cross 
section of selectron, the gaugino mass parameter $M_{1(2)}$  
was determined also with an error of a few \%. 
In their paper, it has also been shown that SUGRA GUT relations such 
as $m_{\tilde e}= m_{\tilde{\mu}}$ and 
 $M_1/M_2= \frac{5}{3}\tan^2\theta_W$ would be checked with 
comparable precision.
 
No analysis in this direction has been made for 
$\st$ production previously. This is because,  for one thing,
 the mode 
is not easy to analyze as the $\tau$ leptons which 
arise from the decay  $\sti\rightarrow\tau\chi^0_i$ 
further decay inside the detector, thus  the kinematics 
is not easy compared to the modes previously studied. 
However, as has been pointed out in \cite{5}, the fact 
that $\tau$ lepton decays further gives 
an interesting opportunity to measure the polarization 
of the $\tau$ lepton ($\ptau$). The polarization is directly 
related to the value of $\tan\beta$, as we discuss 
below.
 
%
\begin{figure}[htb]
\centerline{
\epsfxsize=10cm 
\epsfbox{fig1.r.epsf}
}
\caption[Fig_1]{\label{Fig_1}
}
\end{figure}

Fig.1 shows the interaction of neutral components of 
gauginos and higgsinos to  $\st$ and $\tau$. The interaction 
is completely fixed by the gauge  and supersymmetry of the model. 
The coupling to the gaugino $\tilde{B }$$ (\tilde{W_3})$ is proportinal to 
the gauge coupling $g_{1(2)}$, while the coupling 
of the $\st$ to the Higgino $\tilde{H}_1^0$ is proportional to 
$Y_{\tau}$. The two interactions are different not only in the couplings, but 
also in the chirality of the (s)fermion.
 The (super-) gauge interaction is 
chirality conserving, while the (super-) Yukawa interaction 
flips the chirality. ( In the figure, the arrows next to the
 $\tilde{\tau}$ and $\tau$ 
line show the direction of chirality.) Thus $\ptau$ 
depends on the ratio 
of the chirality flipping and the conserving interactions.
 
As we mentioned already,  gauginos and higgsinos are not 
 mass eigenstates, but they mix to form  
neutralino mass eigenstates $\chi^0_i$. 
 $\str$ and $\stl$ also mix, thus  
the coupling of the $\sti\tau\chi^0_i$  interaction depends on   
both the stau mixing $\theta_{\tau} $ and the neutralino mixing $N_{ij}$.
However, $\theta_{\tau}$ can be determined independently
by measuring the production cross section 
of $\tilde{\tau}$ precisely, as $\sigma(\epem\rightarrow \st^+\st^-)$
depends on $\theta_{\tau}$ only.\footnote{The measurement of 
$\sigma$ is actually correlated to the mixing of neutralino, 
as the detection efficiency depends on the decay modes into 
$\chi^0_i\tau$, which should be studied carefully. 
The detailed discussion can be found in \cite{5}} 
 
For an illustrative purpose, let's discuss the case 
where $\sti=\str$ . Then $\ptau (\str\rightarrow\chii\tau)$ is expressed, 
in the limit that the final state $\tau$ is relativistic, as follows; 
\be
\ptau(\str\rightarrow\chio\tau)
=\frac{(\sqrt{2}g N_{11}\tw)^2-\left( Y_{\tau} N_{13}\right)^2}
{(\sqrt{2}g N_{11}\tw)^2+\left( Y_{\tau} N_{13}\right)^2}.
\label{e15}\ee  
Here $N_{ij}$ is the neutralino diagonalization matrix 
appearing in Eq.(3) and 
$Y_{\tau}$ is the tau Yuakawa coupling  in Eq.(1).
 
%
\begin{figure}[htb]
\centerline{
\epsfxsize=15cm 
\epsfbox{fig2.r.epsf}
}
\caption[Fig_2]{\label{Fig_2}
}
\end{figure}

In fig. 2 we show contours of constant 
$\ptau (\str\rightarrow \chio \tau)$ in
the $M_1-\mu$ plane for $\tanb=10$ (Fig. 2a) and 
$\tanb=2$ (Fig. 2b). $\ptau$
decreases monotonically as $M_1$ increases for a fixed value 
of $\mu$, as$\chio$ is bino--like ($N_{11}\simeq 1$) 
for $M_1\ll\vert\mu\vert$, while $\chio$ is higgsino--like 
if $M_1\gg\vert\mu\vert$ ($N_{11}\ll 1$). 
One should also notice the strong dependence of  $\ptau$ on 
 $\tan\beta$. This is because the Yukawa coupling of the $\tau$ 
lepton increases  linearly with $\tan\beta$ when $\tan\beta$ is large.
 
We learned that the measurement of $\ptau$ gives us a constraint 
to a combination of neutralino mixing $N_{ij}$ and $\tan\beta$. 
Other sparticle productions also carry  information about 
the neutralino sector. However as $\tanb$ becomes larger, the neutralino 
and chargino mass and mixing matrix become less and less dependent on 
$\tanb$.
This is because the off-diagnal elements of 
the mass matrices of the neutralinos and charginos become  insensitive 
to $\tanb$ once  $\tan\beta>10$, as $\cos\beta\sim 0$ and $\sin\beta\sim 1$. 

\begin{figure}[htb]
\centerline{
\epsfxsize=7cm 
\epsfbox{fig3.r.epsf}
}
\caption[Fig_3]{\label{Fig_3}
}
\end{figure}
%
%
 
\begin{figure}[htb]
\centerline{
\epsfxsize=7cm 
\epsfbox{fig4a.r.epsf}
\hskip 0.5cm
\epsfxsize=7cm 
\epsfbox{fig4b.r.epsf}
}
\vskip 0.5cm
\centerline{
\epsfxsize=7cm 
\epsfbox{fig4c.r.epsf}
\hskip 0.5cm
 \epsfxsize=7cm 
\epsfbox{fig4d.r.epsf}
}
\caption[Fig_4]{\label{Fig_4}
}
\end{figure}

To demonstrate this, we show various quantities  in Fig.4 a)-d) 
fixing $\mchi=100$ GeV, and varying $M_1$ and $\tanb$. The 
reasons to take such a parametrization are following:
According to the investigation of the production of  
$\tilde{e_R}^+\tilde{e}_R^-$$(\tilde{\mu}_R^+ \tilde{\mu}_R^-)$ 
and its subsequent decay to 
$e^{\pm}(\mu ^{\pm})\chi^0_1$ in Ref\cite{3},  
one can determine the neutralino 
mass with an error of a few percent from the distribution of the final state
leptons. The same 
analysis is  also possible for the chargino pair production 
and decay, so it would be almost certain that we get 
some idea of the lightest neutralino mass $\mchi$ once 
SUSY particle productions are observed at an $\epem$ collider. The 
corresponding curves that satisfy the constraint were shown in Fig. 3 
on $M_1$ --$\mu$ plain, assuming $\mchi=100$ GeV(no error). 
Here the  solid curve corresponds to $\tan\beta=1.5$ and the dotted curve 
corresponds to  $\tan\beta=30$.\footnote{We assumed GUT relation to the 
gaugino masses} With  the mass constraint, one can 
specify the parameter space of the neutralino sector 
by $M_1$ and $\tan\beta$, up to 
 twofold ambiguity of positive and negative $\mu$ solutions for 
the large $M_1$ and $\tan\beta$ region, 
or up to threefold ambiguity (2 solutions in the negative 
$\mu$ region and one solution 
in the positive $\mu$ region) for the small values of $M_1$ and $\tan\beta$. 
We take the positive $\mu$ solution throughout the plots 
Fig. 4 a)-d) to avoid 
too many lines (sometimes quite close each other) 
appearing  on the same plot.\footnote{The ambiguity of $\mu$ might be removed
by other experiment, such us $Br(b\rightarrow s \gamma)$ \cite{9}} 
 
Fig. 4a and fig. 4b show the contours of the mass differences
 a) $\mchip-\mchi$ and b) $\mchii-\mchi$. One can see the mass 
difference depends very 
mildly on $\tan\beta$ once $\tan\beta>5$. The tendency is also same for the  
negative $\mu$  solution, though the mass differences decreases as $\tan\beta$ 
becomes smaller for the negative $\mu$ solution. 
The dependence on $\tan\beta$ is even  milder  for 
$\sigma(\epem\rightarrow \tilde{e}_R^+\tilde{e}_R^-)$ 
as can be seen in Fig 4c. The production proceeds 
through the s-channel exchange of gauge bosons and 
the t-channel exchange of neutralinos, where 
the dependence on  $M_1$ comes in. The $\tilde{e}_R$ production 
cross section turns out to be the best quantity to 
fix $M_1$, free from the uncertainty of the value 
of $\tan\beta$. 
 
Fig. 4d) is the contour plot of constant $\ptau$
($\str\rightarrow \chi^0_1\tau$). 
The plot looks 
totally different from Fig 4a-c). If $M_1$ is not 
too close to 100 GeV, the polarization 
depends on $\tanb$ sensitively for the parameters 
shown in the figure.
If one knows $M_1$ precisely from the production cross 
section of $\tilde{e}_R$ or from other processes, 
one can extract $\tanb$ by further measuring 
 $\ptau$$(\str\rightarrow \chi^0_1\tau)$. Notice that the sensitivity 
is better in the region $\tan\beta>5$, complimentary 
to the information from the mass differences of --inos.
(See Fig 4a and 4b).
 
For  most of the parameter space, $\mchi$, $\mchii$, $\mchip$ 
are very close to each other, thus the decay mode into those --inos 
are always open. The determination of the branching ratios 
constrain the model parameters even further.
 
Notice the decay into the lightest neutralino may not be 
the dominant decay mode. If the decay modes into 
the gaugino like --ino are open, $\st$ decays  dominantly into it,
even if the higgsino--like --ino 
is lighter than the gaugino--like --ino. 
As the gaugino coupling is insensitive to $\tan\beta$, 
we will not be able to determine $\tan\beta$ 
in such a case.
 
\newpage
 
\section*{4)Measurement of $\ptau$} 
 
The measurement of $\ptau$ 
would be carried out through the 
energy distribution of decay products from the polarized $\tau$ lepton.
The $\tau$ lepton 
 decays into $A\nu_\tau$ where $A=e \nu, \pi,\rho,a_1...$ . The decay
distributions of the $\tau$ decay products depend on the polarization of the
parent $\tau$ lepton\cite{9}. In particular, for each decay channel the momentum
distribution of the decay products ($\pi^-$, $\rho\rightarrow \pi^-\pi^0$,...)
differs significantly depending on whether their parent is $\tau_R^-$($h=1/2$)
or $\tau_L^- $($h=-1/2$). If the $\tau$ lepton is relativistic, $P_{\tau}$ can
then be determined from the energy distribution of the decay products \cite{7}.
Notice  that the $\tau$ lepton from a $\st$ decay also has
 some energy distribution
which depends on $\msti$ and $\mchi$, thus 
the energy spectrum of the final decay products depends 
 on $\msti$, $\mchi$ and $\ptau$. 
\noindent
The three quantity, in principle, can be determined 
from the energy distribution only by fitting 
the energy spectrum, but this is experimentally rather challenging. 
The situation can be improved by using information from other sources:
$\mchi$ from, for instance,
selectron production and decay, and $\msti$ from a threshold scan.
 
\begin{figure}[htb]
\centerline{
\epsfxsize=7cm 
\epsfbox{fig5.r.epsf}
}
\caption[Fig_5]{\label{Fig_5}
}
\end{figure}

In fig. 5  we have shown the normalized energy distribution  
of the $\pi$ ($y=E_{\pi}/E_{\rm beam}$) from 
the cascade decay of $\sti$. 
For the plot we took  $\ptau=\pm 1 , 0$, 
$\msti=150$ GeV, $\mchi=100$ GeV and $\sqrt{s}=500$ GeV. 
The spectrum is considerably harder (softer) for 
$\tau_{R(L)}$. The upper end of the energy distribution ($y_{\rm max}$) 
is the maximum energy of the $\tau$ lepton arising 
from  $\st$, while the minimum energy of the $\tau$ lepton 
corresponds to the peak of the  energy distribution for $\ptau=1$
($y_{\rm \overline{min}}$).
For the parameter used in Fig 5, 52\%(90\%) of the events 
are in $y>y_{\rm\overline{min}}$ region for $\ptau=-1(1)$, 
while 4\% (22\%) of the 
events go above $(y_{\rm \overline{min}}+y_{\rm max})/2$.
 
$\ptau$ can also be measured independently by studying the distributions of
the difference of the energy between decay pions from $\rho$ and $a_1$.
The $\tau\rightarrow\rho\nu$ mode has a branching ratio of about 23\% and 
the $\tau\rightarrow a_1\nu$ mode has a branching ratio of about 15\%. 
$\tau_{L(R)}$ decays dominantly into the longitudinal (transverse) 
element of $\rho$ or $a_1$, and they tend to get most of the $\tau$ energy. 
Then, a transversely polarized $\rho$ favors equal splitting of the $\rho$
energy between the two decay pions, whereas a longitudinally polarized $\rho$
leads to a large difference of the $\pi^-$ and $\pi^0$ energies. For $a_{1T}$,
all three pions have a tendency to share equally the energy of $a_1$. On the
other hand, $a_{1L}$ again favors configurations in which one or two of the
 pions are soft. A detailed discussion of the 
energy distribution may be found in
\cite{7}. 
 
Monte Carlo simulations of the determination of the $\tau$ lepton polarization
are in progress \cite{17}. For this purpose, we have developed a Monte Carlo
event generator of the signal process ($e^+e^-
\rightarrow\tilde{\tau}^+\tilde{\tau}^-$). The generator takes into account
the polarization of decaying taus, by using the TAUOLA2.4 program
package\cite{8}. This package is implemented together with the JETSET7.3
program in a single program module dealing with the hadronization step and is
being used for both the signal and background processes. The generated events
are processed through the standard JLC detector simulator whose parameters can
be fond in \cite{REFDETECTOR}. 
 
As an example to demonstrate the effects of the tau polarization in
the stau decays, we show in Fig.6 the momentum fraction
distributions of $\pi$'s from $a_1$ decays after
a set of cuts to select 1-3 topology. We took $\mst=141.9$ GeV, 
$\mchi=117.8$ GeV and compare the energy distribution for 
$\ptau=1$ and 0.
 
\begin{figure}[htb]
\centerline{
\epsfxsize=7cm 
\epsfbox{fig6.epsf}
}
\caption[Fig_6]{\label{Fig_6}
}
\end{figure}

The original sample contains 5k $\tilde{\tau}$-pair events
in which one tau was forced to decay into $\pi \nu_\tau$
and the other into $a_1 \nu_\tau$.
The background processes such as
$W^+W^-$, $ZZ$, $ZH$, $\gamma\gamma$, etc have not been included yet.
 
 
Studies of the other decay modes are ongoing together with the
background processes.
 
In summary, the measurement of $\ptau(\st\rightarrow\tau\chii)$
could give unique information about $\tan\beta$, combined with the
the information from other measurements such as $\mchii-\mchi$ or 
$\sigma(\epem\rightarrow \tilde{e}^+\tilde{e}^-)$. In some sense 
Fig. 4 might be regarded as key maps of the parameter space of 
MSSM. After years of running of JLC, we may  put our 
finger on a point of the parameter space by the precise measurement of 
event signatures of sparticle productions.
 
\section*{Acknowledgment}
I thank K. Fujii and T. Tsukamoto for collaborations 
which this talk is based on. I would also like to thank M. Drees
for careful reading of the manuscript. This 
work is supported in part by the Grant-in-aid for 
Scientific Research from the Ministry of Education, Science and 
Culture of Japan (06740236).

\end{document}
\bye